\documentclass[useAMS,usenatbib,12pt,psfig]{mn2e} %,referee
\usepackage{color}
\usepackage{graphicx}
\usepackage{txfonts}

\def\hi{H\,{\sc i}}

\def\mh1{$M_{\rm H_{I}}$}

\def\kms{km~s$^{-1}$}

\title[H{\sc  i} debris of Arp 181]{Unusual  displacement of H{\sc i} due to tidal interaction in Arp 181}
\author [Sengupta {\it{et al.}}]{ Chandreyee Sengupta,$^{1,2}$\thanks{e-mail:sengupta@iaa.es(CS), dwaraka@rri.res.in(KSD), djs@ncra.tifr.res.in(DJS),  tom@iaa.es(TS)}  K. S. Dwarakanath$^{3}$, D.J. Saikia$^{4,5}$ and T. C. Scott$^{1}$  \\\\
$^{1}$ Instituto de Astrofısica de Andalucıa (IAA/CSIC), 18080 Granada, Spain \\
$^{2}$ Calar--Alto Observatory, Centro Astron\'omico Hispano Alem\'an, C/Jes\'us Durb\'an Rem\'on, 2-2 04004 Almeria, Spain \\
$^{3}$ Raman Research Institute, Bangalore 560 080, India \\
$^{4}$ National Centre for Radio Astrophysics, Tata Institute of Fundamental Research, Pune 411 007, India \\
$^{5}$ Cotton College State University, Panbazar, Guwahati 781 001, India \\
}

\begin{document}

\date{Received  ; accepted  }
\date{}
\pagerange{\pageref{firstpage}--\pageref{lastpage}} \pubyear{}

\maketitle

\label{firstpage}

\begin{abstract}
We present results from GMRT H{\sc i} 21 cm line observations of the interacting galaxy pair Arp 181 (NGC\,3212 and NGC\,3215) at z =0.032. We find almost all of the detected H{\sc i} (90\%) displaced well beyond the optical disks of the pair with the highest density  H{\sc i} located  $\sim$70 kpc west of the pair. An H{\sc i} bridge \textcolor{black}{extending} between the optical pair and the bulk of the H{\sc i} \textcolor{black}{ together with  their H{\sc i} deficiencies  provide} strong evidence that \textcolor{black}{the interaction between the pair has removed most of their H{\sc i}  to THE current projected position.}  H{\sc i} to the west of the pair has two approximately \textcolor{black}{equal} intensity peaks.  The  H{\sc i} intensity maximum furthest to the west coincides with a small \textcolor{black}{spiral} companion 
SDSS J102726.32+794911.9 which shows enhanced mid--infrared (Spitzer),  UV (GALEX)  and H$\alpha$ emission indicating intense star forming activity.  The H{\sc i} intensity maximum close to the Arp 181 pair, coincides with a diffuse optical cloud detected in UV (GALEX)  at the end of the stellar and H{\sc i}  tidal tails originating at NGC\,3212 \textcolor{black}{and}, previously proposed \textcolor{black}{to be } a tidal  dwarf galaxy in formation. Future sensitive H{\sc i} surveys by telescopes like ASKAP
should prove to be powerful tools for identifying tidal dwarfs at moderate
to large redshifts to explore in detail the evolution of dwarf galaxies in
the Universe. 
\end{abstract}


%\begin{table*} 
%\caption{GMRT observations }
%\begin{tabular}{l l l l l l l rrr}
%\hline
%Frequency & Observation  & Phase      & Phase cal    &  $\tau$ & Bandwidth &rms (per channel  & \multicolumn{3}{c}{Beam size} \\ 
%          & date         & calibrator &  flux        &         &           &for 21-cm line)   &  maj & min & PA                \\
%          &              &            &density (Jy)  & (hr)    & (MHz)     &(mJy beam$^{-1}$) &($^{\prime\prime}$)&($^{\prime\prime}$)
%&($^\circ$)\\ 
%\hline
% 21-cm line & 2008 Apr 30 &J1609+266 &  5.0 & 7   & 8 &  0.5 &   9.0 &  9.0 &      \\
%            &             &          &      &     &   &  0.9 &  29.0 & 21.0 & 119  \\
%            &             &          &      &     &   &  1.0 &  41.0 & 35.0 & 126   \\
%  1403 MHz  &             &          &      &     &   &  0.17&   3.5 & 2.2 & 40  \\
%\hline

%\end{tabular}
%\end{table*}

\section{Introduction}
\label{intro}
Galaxy mergers and interactions  play a key role in cosmological evolution, affecting galaxy morphology, the star-formation history by triggering intense bursts of star formation  \citep{BH1992}, and active galactic nuclei (AGN) via the growth of central supermassive black holes  \citep{Springel05}. Besides  spectacular episodes of star formation,  \textcolor{black}{mergers
can also lead to a significant amount of the progenitor material being temporarily  ejected to large distances in the form of tidal tails and debris}. Interacting galaxies, in relatively  isolated pairs or galaxy groups\textcolor{black}{,}  containing at least one gas rich member,  are  ideally suited \textcolor{black}{to  study  the ways in which  tidally stripped gas evolves, e.g. formation of bridges and tails.} Where such interactions take  place in a group, the  stripped gas evolves under the influence of the intra-group medium (IGM)  while at the same time contributing to it. Studying the  stripped gas can  reveal information about gas dynamics,  star formation, IGM enrichment and the formation of tidal dwarf galaxies (TDGs).

Star formation beyond the galaxy disks and the principles governing it have attracted a lot of recent attention  \citep{Sun2010} and the advent of ultraviolet (UV) and mid-infrared (MIR) telescopes like  GALEX and  Spitzer have enabled major advances in such studies. \textcolor{black}{There were reports that some tidal bridges, tails and debris containing large amounts of H{\sc i} also have  blue optical counterparts, although it is not always clear whether the blue colour arises from young stars removed from the interacting galaxies  or in--situ} star formation from tidally stripped gas. \textcolor{black}{Recent high-resolution MIR and UV observations of  some of these blue counterparts show that they are recently formed stellar clumps}  \citep{Smith, Hibbard, Neff}.  Atomic and molecular gas observations of such systems provide \textcolor{black}{ information for determining} the conditions under which star formation is triggered in these regions.

\textcolor{black}{ If gas densities and environmental conditions within the tidally ejected material are favourable, self-gravitating bodies with  masses typical of dwarf galaxies may host star forming regions} \citep{Duc,ducmirabel}. These entities known as tidal dwarf galaxies (TDGs) \textcolor{black}{ have almost always been found  within tidal debris. Being born of tidally stripped material TDGs have relatively high metallicities compared to normal dwarfs and a lower dark matter content (This is model dependent, see   Barnes \& Hernquist 1992a, Bournaud et al. 2007)    as well as  hosting a mix of old and young stellar populations} \citep{duc2011rev}.  
%velocity dispersions. 

%Typical  H{\sc i} masses of TDGs \textcolor{black}{are a} few times 10$^{8}$ M$_{\odot}$ 
%and stellar masses are \textcolor{black}{ a} few times 
%10$^{9}$ M$_{\odot}$ \citep{Duc}. Besides contributing to the dwarf galaxy population, TDGs provide us with further insights towards understanding
%galaxy evolution \citep{duc2011rev,kaviraj}.

\onecolumn

%\begin{figure}
   %  \hbox{
 %   \centerline{\rotatebox{0}{\includegraphics*[height=7.5in]{figure-comp.eps}} }
%   \centerline{\rotatebox{0}{\includegraphics*[height=4in]{ARP72-MOM0-40ARCSEC-NEW.ps}} }
%    \hspace{-8.2cm}
%    \centerline{\rotatebox{-90}{\includegraphics*[height=3.0in]{Arp 181-system-spectrum.ps}} }
%}     
\begin{figure*}
\centering
\includegraphics[scale=0.75]{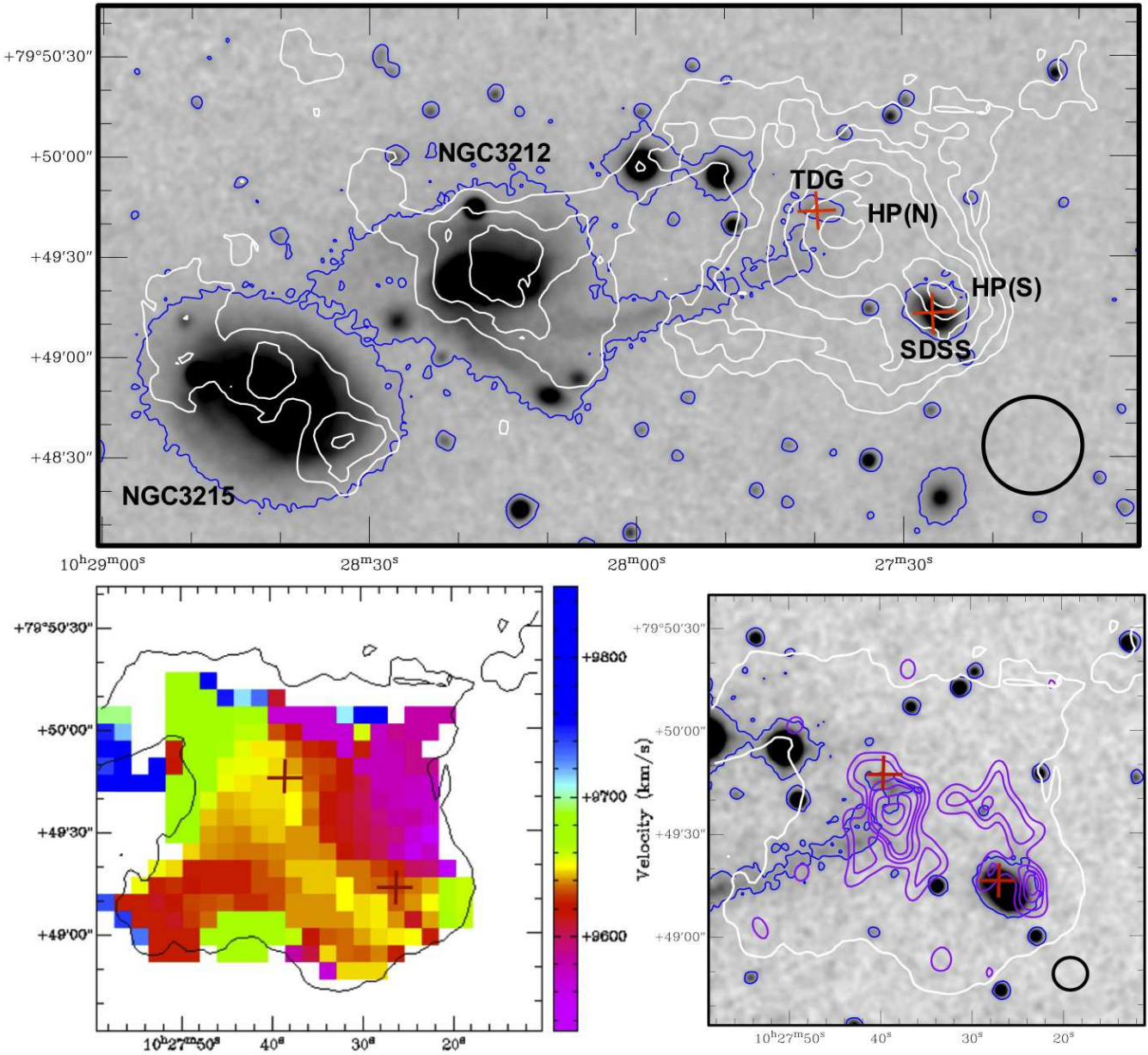}  %[width=6cm]{}
\caption{The top panel shows the low-resolution  H{\sc i} column density contours (white) overlayed on SDSS g -- band image of the Arp 181 system. The column density levels are 10$^{20}$ cm$^{-2}$ $\times$ (1.3, 2.2, 3.3, 4.4, 5.5, 6.6). The diffuse object suggested as a TDG is marked with a red cross as is SDSS  J102726.32$+$794911.9 and are labeled TDG and SDSS respectively. The blue contours mark the faint limit for the g--band image and clearly show tidal tail and the blue diffuse object (TDG) at the end of the tail emanating from NGC 3212. The two H{\sc i} peaks HP(N) and HP(S) and the principal galaxies are also labelled. The beam is indicated with the black ellipse. The lower left panel shows the low resolution velocity field of the tidal debris in colour. The black contour is the outermost column density contour as in the top panel and to show the area of the sky presented in this image. The lower right panel shows the high-resolution H{\sc i} column density contours of the tidal debris overlayed on the  SDSS--ugr image of region containing the TDG and SDSS J102726.32$+$794911.9 . The column density levels are   10 $^{20}$ cm$^{-2}$ $\times$ (1.9, 3.5, 5.5, 6.8, 8.7, 9.9, 12.9, 14.5). The beam is indicated with black ellipse.  The axes are right ascension and declination in J2000 coordinates. } 

%The middle right panel shows  the high resolution H{\sc i} column density contours of the tidal debris overlayed on the smoothed SDSS--r image. The column density levels are   10 $^{20}$ cm$^{-2}$ $\times$ (1.9, 3.5, 5.5, 6.8, 8.7, 9.9, 12.9, 14.5). The bottom left panel is the velocity field of the tidal debris with the contours plotted on the coloured H{\sc i} image of the debris.  are marked with arrows along with the optical radial velocities of NGC\,3212 and NGC\,3215.}
    \label{fig1}
  \end{figure*}

\twocolumn

\noindent  Due to active ongoing star formation they normally have characteristic blue colours and high \textcolor{black}{stellar} velocity
 dispersions. Typical  H{\sc i} masses 
of TDGs \textcolor{black}{are a} few times 10$^{8}$ M$_{\odot}$ and stellar masses are \textcolor{black}{ a} few times 
10$^{9}$ M$_{\odot}$ \citep{Duc}. Besides contributing to the dwarf galaxy population, TDGs provide us with further insights towards understanding
galaxy evolution \citep{duc2011rev,kaviraj}.

In addition to tidally stripped gas and  galactic winds \citep{Ryan-Weber}, in situ star formation in tidal debris can be  an important source   of IGM enrichment. \cite{Ryan-Weber} estimate that \textcolor{black}{a} star-formation rate (SFR) as low as 1.5$\times$10$^{-3}$ M$_{\odot}$ yr$^{-1}$ maintained for 1 Gyr, can raise  the IGM metallicity  by $\sim$1$\times$10$^{-3}$ solar. This value compares well with the ``metallicity floor" $\sim$1.4$\times$10$^{-3}$ solar in the damped Lyman alpha (DLA) gas, observed over a redshift range of 0.5 to 5 \citep{Prochaska}.

 \textcolor{black}{H{\sc i} 21--cm neutral hydrogen observations are a powerful tool for probing the evolution of  tidal debris and identifying TDG  candidates.} In this Letter we present one of the most spectacular cases of the impact on H{\sc i} of a tidal interaction between a pair of spirals, NGC 3212 and NGC 3215, which make up the Arp 181 system and possibly a third small galaxy. NGC 3212 has been classified as an `S?' galaxy at a redshift of 0.0324, while NGC 3215 has been classified as an `SB?' galaxy at a redshift of 0.0316 \citep{rc3}. We adopt a distance of 130 Mpc (assuming H$_{\circ}$=75) for the system and at this distance 1 arcmin translates to  $\sim$ 39.4 kpc.

%  Arp 181 (NGC3215 and NGC 3212), with}  $\ge$ 90 \% of the H{\sc i} is seen in
%two clumps separated from the galaxy pair by $\sim$70 kpc. One of these \textcolor{black}{clumps}
%is associated with a \textcolor{black}{TDG}, demonstrating that sensitive H{\sc i} surveys by telescopes such as ASKAP and in future, the SKA, \textcolor{black}{should} prove to be very powerful tools of identifying tidal dwarfs at
%moderate and large redshifts to explore missing baryons and the evolution
%of our Universe.

%\textcolor{red}{J2000 coordinates are used throughout this letter.}

\section {Observations}
\label{obs}
H{\sc i}  in Arp 181 was observed with the Giant  Metrewave Radio Telescope {\it (GMRT)} on December 31st, 2008. \textcolor{black}{Further} details of the observations  \textcolor{black}{are} given in Table \ref{table1}. The baseband bandwidth used was 8 MHz for the H{\sc i} 21-cm  line observations giving a velocity resolution of $\sim$13 km~s$^{-1}$. To \textcolor{black}{improve the}  signal to noise ratio the velocity resolution was smoothed to $\sim$26 km~s$^{-1}$ \textcolor{black}{to produce the \hi\ maps presented in this Letter}. 

The  GMRT data were reduced using Astronomical Image Processing System ({\tt AIPS}) software package. 
Bad \textcolor{black}{data  from} malfunctioning antennas, antennas with abnormally low gain and/or  radio frequency interference (RFI)
were flagged.   The flux densities are on the scale
of \cite{baar}, with flux density uncertainties of \textcolor{black}{$\sim$5\%.  The} \hi\ cubes were produced following continuum subtraction in the uv domain using the {\tt AIPS} tasks `UVSUB' and `UVLIN'. 
The task `IMAGR'  was used to obtain   the final cleaned  \hi\ image cubes. From these cubes the integrated \hi\ and  \hi\  velocity field maps were extracted using the AIPS task `MOMNT' (USING A 3$\sigma$ CUT-OFF). To analyse the \hi\  structures, we produced image cubes of different resolutions by tapering the data with different uv limits AND USING UNIFORM WEIGHTS FOR THE HIGH RESOLUTION MAPS.

\begin{table}
\centering
\begin{minipage}{110mm}
\caption{GMRT observation details}
\label{table1}
\begin{tabular}{ll}
\hline
%Frequency & Observation  & Phase      & Phase cal    &  $\tau$    & Bandwidth &rms (per channel  & beam size   \\ 

%\textbf{property}&\textbf{value} &\textbf{ reference } \\ 
%\hline

Frequency & 1376.4 MHz \\
Observation Date &31st December, 2008 \\
%Phase Calibrator & 0410+769  \\
%& 1313+675 \\
primary calibrator&3C286\\ 
Phase Calibrators  & 0410+769 (5.76 Jy)  \\
(flux density)  & 1313+675(2.40 Jy)  \\
Integration time  & 8.0 hrs  \\
BASEBAND BANDWIDTH & 8 MHz \\
NUMBER OF CHANNELS & 128 \\
primary beam & 24\arcmin ~at L BAND  \\
Low resolution beam & 32.3$^{\prime\prime}$ $\times$ 30.9$^{\prime\prime}$  (PA = 40.3 $^{\circ}$) \\
% (major axis X minor axis ) &   \\
%PA  & 40.3  deg \\
High resolution beam & 10.5$^{\prime\prime}$ $\times$ 10.5$^{\prime\prime}$ \\
%(major axis X minor axis ) &  \\
%PA (Low resolution) & 40.3 $^{\circ}$ \\
rms for low-resolution map  & 0.70 mJy beam$^{-1}$  \\
rms for high-resolution map & 0.47 mJy beam$^{-1}$  \\
%\hline
RA (pointing centre)& 10$^{\rm h}$ 28$^{\rm m}$ 28.$^{\rm s}$5  \\
DEC (pointing centre)& +79$^\circ$ 49$^\prime$ 05.0$^{\prime\prime}$\\
%Spatial scale&$\sim$39.4 kpc/arcmin\\
%Morphology&NGC\,3212 (S), NGC\,3215 (SB)\\
%Distance &130 Mpc\\

\hline
\end{tabular}
\end{minipage}
\end{table}

\section {Observational Results} 
\label{results}
\textcolor{black}{ Figure \ref{fig1} (upper panel) shows the integrated H{\sc i} low-resolution emission (white contours) in Arp 181, overlayed on a smoothed ($\sim$ 2$^{\prime\prime}$) Sloan Digital Sky Survey (SDSS) g--band  image.  \textcolor{black}{NGC 3212}, NGC 3215, the tidal bridge, the two \hi\ intensity peaks in the  western \hi\ mass (the main concentration of \hi\ west of the principal pair) and the two optical entities associated with the western \hi\ mass are  labelled in the image. The blue contour marks the approximate faint  limit for the SDSS  g--band emission.}   The lower left panel of Figure  \ref{fig1} shows the velocity field of the western \hi\ mass.  \textcolor{black}{The colour scheme in velocity field has been used to  highlight  the velocity gradient within the \hi\ mass. } The lower right panel of \textcolor{black}{ Figure \ref{fig1}} shows the high-resolution H{\sc i} image (contours) of a zoom in to the region where most of the H{\sc i} mass in the system \textcolor{black}{was} detected.

Figure  \ref{fig1} (upper panel),  shows that almost the entire H{\sc i} mass of the system is detected well beyond the optical disks of   NGC\,3212 and NGC\,3215 and is  connected to  NGC\,3212 by a low density H{\sc i} bridge.  The H{\sc i} \textcolor{black}{bridge is} not coincident with the optical tidal tail but runs parallel \textcolor{black}{and to  the north of it}.  The western \hi\ mass has two distinct intensity maxima at positions  10$^{\rm h}$ 27$^{\rm m}$ 37.$^{\rm s}$1, 79$^\circ$ 49$^\prime$ 36.7$^{\prime\prime}$ and 10$^{\rm h}$27$^{\rm m}$ 25.$^{\rm s}$1,  79$^\circ$ 49$^\prime$ 16.62$^{\prime\prime}$, hereafter referred to as HP(N) and HP(S) respectively.  HP(S) approximately coincides with a small blue spiral 
SDSS J102726.32+794911.9 \textcolor{black}{which has} no published redshift. however information in   $http://www.etsu.edu/physics/bsmith/research/sg/sara_sg.html$ confirms this galaxy is at similar redshift to the   ARP 181 system.   HP(N) \textcolor{black}{ is close to a diffuse blue object at the projected end of the optical tidal tail emanating } from NGC 3212 (Figure \ref{fig1}). 

%\begin{figure}
%\centering
%\includegraphics[scale=1.05]{arp181_hres.eps}  %[width=6cm]{}
%\caption{ High resolution H{\sc i} column density contours of the tidal debris overlayed on the  SDSS--ugr image of region containing the TDG and SDSS J102726.32$+$794911.9 . The column density levels are   10 $^{20}$ cm$^{-2}$ $\times$ (1.9, 3.5, 5.5, 6.8, 8.7, 9.9, 12.9, 14.5). The beam sise and orientation  is indicated with blue ellipse.}
%    \label{fig0}
%  \end{figure}

The total H{\sc i} flux density recovered from this system from the GMRT observations is  3.3 $\pm$ 0.2 Jy~\kms\ \textcolor{black}{  compared } to the single dish value of 2.3 $\pm$ 1.5 Jy~\kms \citep{sd}. Using \textcolor{black}{ a GMRT integrated} H{\sc i} map, the H{\sc i} content of individual galaxies and the western mass was estimated. The two galaxies, NGC\,3212 and NGC\,3215  contain \textcolor{black}{ \hi\ of}  2.7$\times$ 10$^{9}$ M$_{\odot}$ and 7.9$\times$ 10$^{8}$ M$_{\odot}$ respectively \textcolor{black}{while}  the western H{\sc i} mass along with the tidal bridge \textcolor{black}{is  9.5$\times$ 10$^{9}$ M$_{\odot}$} The uncertainty on \textcolor{black}{these} estimates are $\sim$10\%.  The velocity field of the western H{\sc i} mass  shows no signs of regular rotation at this velocity resolution. However there is a gradient from both east and west, converging near the positions of the \hi\ intensity maxima, especially HP(N).

\section {Discussion}
\label{dis}
 Within the limits of current spectroscopic data Arp 181 appears to be a  fairly isolated system. The nearest similar size neighbour is CGCG 350-053  projected at an angular distance of $\sim$30.4$^{\prime}$  (1.4 Mpc) with a velocity separation of $\sim$ 2000 \kms. There are several small galaxies  projected close to Arp 181, but they lack spectroscopic data to confirm their distances. Thus it is possible Arp 181 is a galaxy group dominated by NGC\,3212 and NGC\,3215. The system has been previously observed with Spitzer \citep{Smith}, GALEX \citep{Smith2010a}  and in H$\alpha$ ($http://www.etsu.edu/physics/bsmith/research/sg/sara_sg.html$). 

Based on their GALEX and optical observations, \cite{Smith2010b} conclude this system  potentially hosts  one or more TDGs. They suggested the two objects at positions  10$^{\rm h}$ 27$^{\rm m}$ 26.$^{\rm s}$2, 79$^\circ$ 49$^\prime$ 12.3$^{\prime\prime}$ \textcolor{black}{(SDSS J102726.32+794911.9)} and 10$^{\rm h}$ 27$^{\rm m}$ 40.$^{\rm s}$1,  79$^\circ$ 49$^\prime$ 45.3$^{\prime\prime}$ \textcolor{black}{ are TDG candidates.} Their optical spectroscopic observations confirmed the galaxy at position 10$^{\rm h}$ 27$^{\rm m}$ 26.$^{\rm s}$2,  79$^\circ$ 49$^\prime$ 12.3$^{\prime\prime}$  \textcolor{black}{(SDSS J102726.32+794911.9) is associated with the }system. This galaxy appears to have spiral features in the SDSS images, and  \textcolor{black}{is} extremely blue in NUV -- g.  According to  \cite{Smith2010b},  this may be a dwarf galaxy which  \textcolor{black}{is} a part of the group or a recently detached TDG.  The object at position  10$^{\rm h}$ 27$^{\rm m}$ 40.$^{\rm s}$1,  79$^\circ$ 49$^\prime$ 45.3$^{\prime\prime}$,is their second TDG candidate, although its association with the Arp 181 system  remains to be  confirmed  spectroscopically \citep{Smith2010b}. The \textcolor{black}{ GALEX observations at the position  of this second candidate showed it to be  a diffuse clump of UV emission at the tip of the optical tidal tail emanating from NGC 3212. Unlike SDSS J102726.32+794911.9, which is bright in Spitzer and SDSS images, this clump is prominent only in UV. In Figure  \ref{fig1} (upper panel) this second TDG candidate is faintly visible at the tip of the tidal tail emanating from NGC 3212 and is near  to HP(N).}

Our observations reveal \textcolor{black}{ only moderate \hi\ emission from the interacting pair, NGC\,3212 and NGC\,3215, with the bulk of the H{\sc i} emission detected in the western \hi\ mass well beyond the optical disks of the Arp 181 pair. The two dwarf systems mentioned in \cite{Smith2010b} are both projected on the western \hi\ mass. Such a large amount  of H{\sc i}  ($\sim$  9.5$\times$ 10$^{9}$ M$_{\odot}$) beyond the principal galaxies' optical disks and associated  with only small optical counterparts or star forming regions is highly unusual. Two likely scenarios to account for this unusual \hi\ distribution are discussed below.} 

Tidal interaction between NGC 3212, NGC 3215 and SDSS J102726.32+794911.9 seems to be a likely \textcolor{black}{ source of the western \hi\ mass} as there is no nearby neighbour of similar size in similar redshift range within $\sim$ 3Mpc radius. Also both the Arp 181 \textcolor{black}{pair galaxies are} severely H{\sc i} deficient indicating \textcolor{black}{ \hi\ has been removed} from the galaxies. \textcolor{black}{A method} to determine gas loss from a spiral is to compare its H{\sc i} surface density \textcolor{black}{to that of a sample of field spirals of the same morphological type. The parameter used to  estimate the \hi\ }  deficiency is log${{\frac{M_{H_{I}}}{D_{l}^{2}}}}$ \citep{haynes}, where where $M_{H_{I}}$ is the total H{\sc i} mass of a galaxy and $D_{l}$ is the optical major isophotal diameter (in kpc) measured at or reduced to a surface brightness level m$_B$ = 25.0 mag/arcsec$^{2}$. While \cite{haynes} used the UGC major diameters, we have used the RC3 diameters and used the modified values from \cite{sengupta06}. The expected value of this parameter for an early type spiral (Sb) is 6.91 $\pm$ 0.26. Since morphological sub-classifications are not available for NGC 3212 and NGC 3215, we compared these galaxies against Sb type spirals. The H{\sc i} surface density values for NGC 3212 and NGC 3215  are   \textcolor{black}{5.90 and 5.66, respectively, indicating they are }  H{\sc i} deficient by a factor of 10. \textcolor{black}{Because of the morphological type, diameter and  H{\sc i}  mass uncertainties, the values for the H{\sc i} deficiencies are themselves highly uncertain. Despite this it remains clear that taken together the pair is highly \hi\ deficient. This lost \hi\ is likely to be  be a major source of the \hi\ in the western \hi\ mass.} The tidal bridge connecting NGC\,3212 to the \textcolor{black}{western \hi\ mass}  also supports  that this H{\sc i} mass is a tidal debris from the Arp 181 interaction. The small dwarf system, SDSS J102726.32+794911.9, may also be  participating in the interaction. The high-resolution H{\sc i} image (lower right of Figure \ref{fig1}) shows the HP(S) to be offset by $\sim$ 5$^{\prime\prime}$ (a beam) from the optical centre of SDSS J102726.32+794911.9 and its H{\sc i} extends in the direction of the candidate TDG. However, given its size, the H{\sc i} mass contributed by this system \textcolor{black}{is likely to be a small fraction of the western  H{\sc i} mass.}

Another scenario \textcolor{black}{for the origin of the western \hi\ mass could be that of SDSS J102726.32+794911.9 being a low surface brightness (LSB) galaxy and the western H{\sc i} mass is at least partially the galaxy's \hi\ disk}. Under this scenario HP(S) is \textcolor{black}{ the central region of the} LSB. The SDSS images show distinct spiral pattern in SDSS J102726.32+794911.9 and thus it is possible that only the central bright part of this galaxy is visible. However, two significant \textcolor{black}{observational facts do not support} this scenario. The H{\sc i} column density peak value for the high-resolution map reaches a higher value than the low-resolution map, indicating the \textcolor{black}{ \hi\ } to be compact and clumpy. The peak values and dimensions of the H{\sc i} clumps in the high-resolution H{\sc i} map are similar for HP(N) and HP(S), consistent with both being \textcolor{black}{similar entities}. Also contrary  to what is expected of a big LSB spiral, we find no signs of regular rotation in this \textcolor{black}{ \hi\ } mass. Instead the velocity field of the western \hi\ mass is chaotic and (lower panel  Figure \ref{fig1}) shows an interesting gradient suggesting infall of gas from the eastern and the western sides towards the central region. Thus we find this \textcolor{black}{ \hi\ mass to be behaving} more like  tidal debris than an interacting massive LSB disk. 

%\begin{figure}
%\centering
%\includegraphics[scale=0.42]{ARP181_zoom_hr2_hi.eps}  %[width=6cm]{}
%\caption{ High resolution H{\sc i} column density contours of the tidal debris overlayed on the  SDSS--ugr image of region containing the TDG and SDSS J102726.32$+$794911.9 . The column density levels are   10 $^{20}$ cm$^{-2}$ $\times$ (1.9, 3.5, 5.5, 6.8, 8.7, 9.9, 12.9, 14.5). The beam is indicated by a blue ellipse.}
%    \label{fig0}
%  \end{figure}

 As explained in Section 3, the debris  has two peaks, HP(N) and HP(S), and we confirm these peaks coincide, with small offsets ($\sim$ 5$^{\prime\prime})$, with the two dwarf galaxies described in \cite{Smith2010b}, the TDG and SDSS J102726.32+794911.9 respectively.   From the channel images and the GMRT spectrum, we find \textcolor{black}{the \hi\ velocity of  HP(N)  is 9659 kms$^{-1}$. The high-resolution  \hi\ zoomed in image of HP(N) and HP(S) (Figure \ref{fig1}) shows a diffuse blue object close to HP(N), at the tip of the tidal tail emanating from NGC\,3212.} This is the TDG candidate suggested by \cite{Smith2010b}.  The \textcolor{black}{ \hi\  column density maximum} in the low-resolution ($\sim$ 30$^{\prime\prime}$) map for this object is 6.6$\times$10$^{20}$cm$^{-2}$. At a distance of 130 Mpc, a 30$^{\prime\prime}$ beam samples $\sim$18 kpc. The \hi\  mass of the debris and the H{\sc i} column density values are a good match with those  found  in TDGs \citep{braine,duc1997}. The high-resolution ($\sim$ 10$^{\prime\prime}$) H{\sc i} map  further confirms the H{\sc i} peak to be associated WITH the TDG as well as  H{\sc i} extension to the east towards the \textcolor{black}{optical} tidal tail of NGC\,3212.

HP(S) on the other hand, coincides with the small, very blue galaxy \textcolor{black}{SDSS J102726.32+794911.9, which \cite{Smith2010b} suggested is} either a dwarf galaxy or a recently detached TDG. Our high-resolution H{\sc i} image (lower right of Figure \ref{fig1}) shows an offset ($\sim$ 5$^{\prime\prime}$) between the galaxy and the H{\sc i} peak. From the GMRT spectrum, we find the H{\sc i} \textcolor{black}{velocity of HP(S) to be $\sim$ 9575 kms$^{-1}$, indicating it is} either interacting with the tidal debris \textcolor{black}{ from the Arp pair or  formed from it.} The latter option is unlikely as the galaxy is bright in SDSS z -- band suggesting presence of an old stellar population \textcolor{black}{ and its faint spiral structure would be unusual for} a recently formed TDG.  Unlike the diffuse \textcolor{black}{optical} structure associated with  HP(N), \textcolor{black}{SDSS J102726.32+794911.9 is}  bright in the Spitzer images implying strong star formation activity, possibly triggered by interaction. \textcolor{black}{  Additionally, the offset  between  HP(S) and SDSS J102726.32+794911.9 may  be the result of an ongoing interaction with the \hi\ debris from the Arp 181 pair  and the \hi\ disk of SDSS J102726.32+794911.9, which is itself  part of the Arp 181 system. }

Massive displacement of H{\sc i} from the galaxy disk due to tidal interaction has been witnessed in a few other cases also \citep{n4438, arp143, duc1997}.  Of these, the H{\sc i} distribution of the Arp 181 system very closely resembles \textcolor{black}{that in } Arp 105 \citep{duc1997}. In the Arp 105 system, bulk of the \textcolor{black}{ $\sim$ 6$\times$10$^{9}$ M$_{\odot}$ H{\sc i} was detected at the end of a 100 kpc long optical tidal tail originating from the \hi\ rich spiral NGC 3561A, with the disk of NGC 3561A  containing only $\sim$ 5$\times$10$^{7}$ M$_{\odot}$ of \hi\.~  The bulk of the \hi\ is connected to NGC 3561A by a faint and discontinuous H{\sc i} bridge and is also host to a TDG candidate A105N. The Arp 105 system is known as an example of extreme gas segregation, with the  H{\sc i} and CO in this system having been completely segregated as a result of  interaction between NGC 3561A and NGC 3561B. Molecular gas mass $\sim$  10$^{10}$ M$_{\odot}$  was detected in the NGC 3561A disk with no detection of molecular gas in the tidal debris or anywhere else in the system \citep{duc1997}. Based on Arp 105, it is possible that the disks of  NGC 3215 and NGC 3212 are H{\sc i} deficient but have retained their molecular gas. So molecular gas observations are required for Arp 181 to get a complete picture of the gas distribution of this interesting system. }

\section{Summary and concluding remarks}
\textcolor{black}{ In summary,  Arp 181 appears to be at a particularly interesting stage of its evolution  following the interaction between at least two, possibly three, of its members during which most of the  H{\sc i} in the system has been stripped from the parent galaxies and 9.5$\times$ 10$^{9}$ M$_{\odot}$ of \hi\ tidal debris is now observed $\sim$ 70 kpc west of  the  principal galaxy pair. We find this \hi\ debris to be in in a state of rapid evolution including hosting  at least one tidal dwarf in formation. We confirm the radial velocity of this \hi\  in the region of this TDG is $\sim$ 9659  km s$^{-1}$.  \hi\ debris also contains another small star forming galaxy, its optical and IR properties suggest it to be a pre-existing member of Arp 181.} Arp 181 is a system at moderate redshift $\sim$ 0.032. Future sensitive H{\sc i} surveys by telescopes such as Australian Square Kilometre Array Pathfinder (ASKAP)  \citep{askap1,askap2} \textcolor{black}{should} prove to be very powerful tools for identifying tidal dwarfs at moderate and large redshifts to explore in detail the evolution of dwarf galaxies in the Universe.
%moderate and large redshifts

%  Arp 181 (NGC3215 and NGC 3212), with}  $\ge$ 90 \% of the H{\sc i} is seen in
%two clumps separated from the galaxy pair by $\sim$70 kpc. One of these \textcolor{black}{clumps}
%is associated with a \textcolor{black}{TDG}, demonstrating that sensitive H{\sc i} surveys by telescopes such as ASKAP and in future, the SKA, \textcolor{black}{should} prove to be very powerful tools of identifying tidal dwarfs at
%moderate and large redshifts to explore missing baryons and the evolution
%of our Universe.

\section{Acknowledgments}

We thank the staff of the {\it GMRT} who have made these observations possible. The {\it GMRT} is operated by 
the National Centre for Radio Astrophysics of the Tata Institute of Fundamental Research. We thank the reviewer for his/her useful comments and
suggestions. This research 
has made use of the NASA/IPAC Extragalactic Database (NED) which is operated by the Jet Propulsion Laboratory, 
California Institute of Technology, under contract with the National Aeronautics and Space Administration.

%%%%%%%%%%%%%%%%%%%%%%%%%%%%%%%%%%%%%%%%

%%%%%%%%%%%%%%%%%%%%%%%%%%%%%%%%%%%%%%%%

\end{document}